\documentclass[prl, showpacs, twocolumn]{revtex4}

\def\b{\begin{eqnarray}}
\def\e{\end{eqnarray}}

\def\a{\alpha}
\def\p{\partial}
\def\l{\lambda}

\def\s{\sigma}
\def\d{\delta}
\def\f{\phi}
\def\g{\gamma}
\def\G{\Gamma}
\def\2{\frac{1}{A^2 - \phi}} 
\def\ph{\phantom{\mu}}
\def\phj{\phantom{j}}

\begin{document}

\title{Riemann Tensor of the Ambient Universe, the Dilaton and the Newton's Constant}
\author{Rossen I. Ivanov} 
\affiliation{Institute for Nuclear Research and Nuclear Energy, Bulgarian Academy of Sciences, 
72 Tsarigradsko chaussee, Sofia -- 1784, Bulgaria and \\
School of Electronic Engineering, Dublin City University, Ireland}
\email{rossen.ivanov@dcu.ie}
\author{Emil M. Prodanov}
\affiliation{School of Physical Sciences, Dublin City University, Ireland} 
\email{prodanov@physics.dcu.ie}

\begin{abstract}
We investigate a four-dimensional world, embedded into a five-dimensional spacetime, and find the five-dimensional Riemann tensor via 
generalisation of the Gauss (--Codacci) equations. We then derive the generalised equations of the four-dimensional world and also show 
that the square of the dilaton field is equal to the Newton's constant. We find plausable constant and non-constant solutions for the 
dilaton.

\end{abstract}

\maketitle

Since the pioneering works of Kaluza \cite{kaluza} and Klein \cite{klein}, who unified gravitation with 
electromagnetism, the implications of possible extra dimensions to our world have been under intense 
investigation --- see \cite{acf} for an extensive collection of papers on higher-dimensional unification. 
Jordan \cite{j} and Thiry \cite{th} used the equations of Kaluza-Klein's theory to show that the gravitational constant
can be expressed as a dynamical field. A constant solution for the Newton's constant however is then allowed 
only if the square of the Maxwell electromagnetic tensor vanishes. Here we examine a {\it dual} set up in which this   
problem can be avoided. Based on the original Kaluza's model, we are here considering a general embedding of 
a four-dimensional world into a five-dimensional ambient spacetime. We derive a generalisation of 
Gauss (--Codacci) equations by utilising all degrees of freedom (the entire geometry) of the ambient 
spacetime and we also show how they affect the physics of the four-dimensional world. As a result we find 
a system of equations for the electromagnetic field, gravitational field and dilaton field. One of these 
equations is a plausable generalisation of gauge fixed Maxwell equations in the presence of a dilaton field.
We also show that the square of the dilaton field is equal to (modulo numerical factors) the Newton's 
constant. The gauge freedom of the electromagnetic fields is transferred to a freedom in fixing the dilaton field.
Apart from the constant solution for the dilaton, we give an example for a non-constant solution 
describing time-varying Newton's constant in an expanding universe (see also \cite{dirac}--\cite{scm} and others). We also give a 
general formula for generating different solutions for the dilaton field.

We consider a four-dimensional world M, embedded into a five-dimensional spacetime V (see \cite{emb1}--\cite{adm} and the references 
therein for a detailed discussion on embeddings). 
Let $y^i \; (i = 1, 2, 3, 4) $ denote the coordinates on M and $x^\mu \; (\mu = 1, 2, 3, 4, 5) $ denote the 
coordinates on V. Greek indexes will be related to the five-dimensional spacetime V, while Latin indexes 
will be associated with the four-dimensional spacetime M. Let $\psi(x^\mu) = s = \mbox{const}$ be the 
equation of the hypersuface M. One can alternatively express the parametric equations of M as 
$x^\mu = x^\mu(y^j, s)$ and treating the parameter $s$ as a coordinate, this then represents a coordinate 
transformation with inverse:
\b
y^j = y^j(x^\mu) \, ,  \quad s = \psi(x^\mu) \, .
\e
We assume that this transformation is invertible at each point. This means that the Jacobi matrices of the 
transformation and its inverse have non-vanishing determinants everywhere. Thus, to globally parametrise the
foliated five-dimensional spacetime V, it is sufficient to use the coordinates of the four-dimensional world 
M and the foliation parameter $s$. \\
The vector normal to the surface is:
\b
N_\mu  =  \frac{\p \psi}{\p x^\mu} \, . 
\e
Let us also define:
\b
e^\mu_j & = & \frac{\p x^\mu}{\p y^j} \, , \qquad n^\mu  =  \frac{\p x^\mu}{\p s} \, , 
\qquad E_\mu^i  =  \frac{\p y^i}{\p x^\mu} \, . 
% n^\mu & = & \frac{\p x^\mu}{\p s} \, , \\
% E_\mu^i & = & \frac{\p y^i}{\p x^\mu} \, .
\e
The derivatives are related as follows:
\b
\p_k \, \equiv \, \frac{\p}{\p y^k} & = & e^\mu_k \, \p_\mu \, , \\
\label{s}
\p_\nu \, \equiv \, \frac{\p}{\p x^\nu} & = &  E^k_\nu \, \p_k + N_\nu \, \p_s \, .
\e
Obviously, if we denote $e^\mu_5 = n^\mu$ and $E^5_\mu = N_\mu$, then $(e^\mu_\nu)$ and $(E^\mu_\nu)$ 
will be the Jacobi matrices of the transformation $(x^\mu) \rightarrow (y^i, s)$ and its inverse. Therefore:
\b
\label{ort}
e^\mu_\s E^\s_\nu = \d^\mu_\nu =  E^\mu_\s e^\s_\nu \, .
\e
This orthogonality condition is equivalent to:
\b
\label{ort2}
e^\mu_i E^j_\mu & = & \d^j_i \, , \\
N_\mu n^\mu     & = & 1 \, , \\
E^i_\mu e^\nu_i + N_\mu n^\nu & = & \d^\nu_\mu \, , \\ 
N_\mu \, e^\mu_j & = & 0 \, , \\
\label{ort3} 
n^\mu E^j_\mu  & = & 0 \, .
\e   
(Note that $\d^\mu_\mu = \mbox{dim V} = 5 \mbox{  and } \d^i_i = \mbox{dim M} = 4$.) \\ 
Thus the bases $(e^\mu_j, n^\mu)$ and $(E^j_\mu, N_\mu)$ are dual. They do not depend on the metric of either
spacetime, but only depend on the particular embedding chosen. \\
Let us now introduce a scalar field $\f(y^k, s)$, a vector field $A_i(y^k, s)$ and the metric tensor 
$g_{ij}(y^k, s)$ on M. We further define the metric $G_{\mu \nu}$ of the five-dimensional spacetime V as an 
expansion over the basis vectors $E^i_\mu$ and $N_\mu$: 
\b
\label{G}
G_{\mu \nu} = E^i_\mu E^j_\nu \, g_{ij} + (N_\mu E^i_\nu + N_\nu E^i_\mu) A_i + N_\mu N_\nu \f \, .   
\e
Taking $x^i =  y^i \, , \; x^5 = s = \mbox{const, i.e. } \; e^\mu_i = \d^\mu_i \, , n^\mu = 
\d^\mu_5 \, , E^i_\mu = \d^i_\mu \, , N_\mu = \d_\mu^5 \, $ in (\ref{G}) corresponds to the
original Kaluza's model \cite{kaluza}. Klein's modification \cite{klein} 
$g_{ij} \rightarrow g_{ij} + A_i A_j$, together with the identification of $\f$ as a dilaton is the model 
put forward by Jordan \cite{j} and Thiry \cite{th}. We note that the metric (\ref{G}) has the same form as the inverse of the metric of 
Klein's model, thus the two theories are dual: Kaluza's model corresponds to slicing, while Klein's model corresponds to threading of the
five-dimensional spacetime \cite{boersma}. The case with $A_i = 0$ has also been considered (see, for example, \cite{adm}, \cite{sms} and 
the references therein). \\
The lack of gauge invariance for the fields $A_i$, which we nevertheless will associate with the electromagnetic potentials, in view of the
slicing--threading duality, is compensated by the freedom to fix $\f$. This, as will become claer later is the freedom to fix the dilaton
field. \\ 
Returning to (\ref{G}), we define $g^{ij}$ as the inverse of the metric $g_{ij}$.
Thus $A^i = g^{ij} A_j$ and $A^2 = g^{ij} A_i A_j$. \\    
The inverse $G^{\mu \nu}$ of the metric $G_{\mu \nu}$ on V is then given by:
\b
\label{Ginv}
G^{\mu \nu} & = & h^{ij} e^\mu_i e^\nu_j - \theta A^i (e^\mu_i n^\nu + e^\nu_i n^\mu) 
+ \theta n^\mu n^\nu \, , 
\e
where $\theta = (\f - A^2)^{-1}$ and $h^{ij} = g^{ij} + \theta A^i A^j$. One can easily check that 
$G^{\mu \l} G_{\l \nu} = \d^\mu_\nu$. 
Using the inverse $G^{\mu \nu}$ (\ref{Ginv}) of the metric $G_{\mu \nu}$, we can raise and lower 
five-dimensional indexes to get:
\b
\label{Ncont}
N^\mu & = & G^{\mu \nu} N_\nu \, = \, \theta (n^\mu - A^i e^\mu_i) \, , \\
\label{N2}
N^2 & = & N_\mu N^\mu \, = \,  \theta \, , \\
\label{ncov}
n_\mu & = & G_{\mu \nu} n^\nu \, = \, A_i E_\mu^i + N_\mu \f \, , \\
\label{n2}
n^2 & = & n_\mu n^\mu \, = \, \f \, .  
\e
Note that when $A^i = 0$ and $\f = 1$, then $N^\mu = n^\mu$ as in the ADM approach \cite{adm}. \\
The extrinsic curvature is: 
\b
\label{K}
K_{jl} & = & e^\mu_j e^\nu_l Q_{\mu \nu} \, ,
\e
where $Q_{\mu \nu} = \nabla_\mu N_\nu = \nabla_\nu N_\mu$. \\
Multiplying (\ref{K}) across by $E_\a^j E_\beta^l$ and applying the orthogonality conditions 
(\ref{ort2})--(\ref{ort3}), one easily finds:
\b
Q_{\a \beta} = E_\a^i E_\beta^j K_{ij} + (N_\a E^i_\beta + N_\beta E^i_\a) f_i + N_\a N_\beta \, \chi \, , 
\e
with $\chi = n^\mu n^\nu Q_{\mu \nu} \, $ and
\b 
\label{f}
f_j  =  n^\mu e^\beta_j Q_{\mu \beta} = A^i K_{ij} + \frac{1}{N} \p_j N \, ,
\e
where $N = \sqrt{N_\mu N^\mu}$. \\
The four-dimensional Christoffel symbols 
\b
\g^i_{jk} = \frac{1}{2} g^{il} (\p_k g_{lj} + \p_j g_{lk} - \p_l g_{jk}) 
\e
can then be expressed as:
\b
\label{christoffel}
\g^i_{jl} = (\p_j e^\mu_l + e^\a_j e^\beta_l \G^\mu_{\a \beta}) E^i_\mu - K_{jl} A^i \, , 
\e
where $\G^\mu_{\a \beta}$ are the five-dimensional Christoffel symbols. Further, the four-dimensional 
Riemann curvature tensor
\b
r^i_{\phj jkl} = \p_k \g^i_{jl} - \p_l \g^i_{jk} + \g^m_{jl} \g^i_{mk} - \g^m_{jk} \g^i_{ml} \, , 
\e
using (\ref{christoffel}) and (\ref{ort2})--(\ref{ort3}), becomes:
\b
\label{gauss}
r^i_{\phj jkl} & = & E^i_\a e^\l_j e^\mu_k e^\nu_l R^\a_{\ph \l \mu \nu}  
+ E^i_\a (\nabla_\mu n^\a) (e^\mu_k K_{lj} - e^\mu_l K_{kj}) \nonumber \\
& & \hskip20pt + \nabla_l (A^i K_{kj}) - \nabla_k (A^i K_{lj}) \nonumber \\ 
& & \hskip20pt + A^i A^m (K_{ml} \, K_{kj} - K_{mk} \, K_{lj}) \, ,
\e
where $R^\a_{\ph \l \mu \nu}$ is the five-dimensional Riemann curvature tensor. The above is a 
{\it generalisation} of the Gauss equations. Taking $A^i = 0$ and $\f = 1$, one simply recovers the well 
known Gauss equations (see, for example, \cite{wald}):
\b
r^i_{\phj jkl} & = & E^i_\a e^\l_j e^\mu_k e^\nu_l R^\a_{\ph \l \mu \nu} + 
K^i_k \, K_{lj} - K^i_l \, K_{kj} \, .
\e 
Let us now expand the five-dimensional Riemann curvature tensor over our basis. Using its symmetries we 
can write:
\b
\label{Riemann}
R_{\l \mu \nu \s} & = & E^i_\l E^j_\mu E^k_\nu E^l_\s U_{ijkl} + \nonumber \\
& & + \Bigl[ (N_\l E^j_\mu - N_\mu E^j_\l) E^k_\nu E^l_\s  \nonumber \\
& & \hskip60pt + (N_\nu E^j_\s - N_\s E^j_\nu) E^k_\l E^l_\mu \Bigr] V_{jkl}  \nonumber \\ 
&& + (N_\l E^j_\mu - N_\mu E^j_\l) (N_\nu E^l_\s - N_\s E^l_\nu) W_{jl} \, ,  
\e
where the coefficients in this expansion satisfy:
\b
U_{ijkl} & = & U_{klij} \, \, = \, \,  - U_{ijlk} = - U_{jikl} \, , \\
V_{jkl} & = & - V_{jlk} \, , \, \qquad W_{jl}  =  W_{lj} \, .
\e
Using (\ref{gauss}), one can further find:
\b
\label{u}
U_{ijkl} & = & e_i^\l e_j^\mu e_k^\nu e_l^\s R_{\l \mu \nu \s} \nonumber \\ 
& = & r_{ijkl} - (\pi_{ik} \, \pi_{lj} - \pi_{il} \, \pi_{kj}) \, ,\\
\label{v}
V_{jkl}  & = & n^\l e^\mu_j e^\nu_k e^\s_l R_{\l \mu \nu \s} \nonumber \\
& = & A^i U_{ijkl} - \frac{1}{N} (\nabla_k \, \pi_{lj} - \nabla_l \, \pi_{kj}) \, ,    
\e
where $\pi_{jl} = \frac{1}{N} K_{jl}$. \\
Finding the remaining tensor $W_{jl}$ is more complicated. One can easily see that
\b
\label{W}
W_{jl} = n^\mu n^\s e^\l_j e^\nu_l R_{\l \mu \nu \s} = S_{jl} + A^k V_{lkj} \, ,
\e
where
\b
\label{S}
S_{jl} = \frac{1}{N^2} n^\s e^\l_j e^\nu_l (\nabla_\l Q_{\s \l} - \nabla_\s Q_{\nu \l}) \, .
\e
In the above we identify derivatives in the direction of $n^\s$. To handle this type of terms, we will have 
to explicitly invoke the dependence on $s$. \\
Firstly, in virtue of (\ref{s}), we get the following expression for the extrinsic curvature (\ref{K}):
\b
\label{KK}
K_{jl} & = & - \frac{N^2}{2} (\nabla_j A_l + \nabla_l A_j) + \frac{N^2}{2} \p_s g_{jl} \nonumber \\ 
& & - \frac{1}{2} N^\l (g_{kl} \p_j E^k_\l + g_{kj} \p_l E^k_\l + A_l \p_j N_\l + A_j \p_l N_\l) \nonumber \\
& & + \frac{N^2}{2} G_{\mu \nu} \Bigl[ A^k \p_k (e^\mu_j e^\nu_l) - \p_s (e^\mu_j e^\nu_l) \Bigr] \, .
\e
Using (\ref{f}), (\ref{W}), (\ref{S}) and (\ref{KK}), it follows that
\b
\label{tuka}
W_{jl} \!\! & = \!\! & 
\frac{1}{N^3} \nabla_j \nabla_l N - \frac{2}{N^4} (\p_j N)(\p_l N) + A^i A^k U_{ijkl} \nonumber \\ 
& & + \frac{1}{N} \Bigl[ \nabla_j \, (A^k \pi_{kl}) + \nabla_l \, (A^k \pi_{kj})\Bigr]
 - \frac{1}{N} A^k \nabla_k \, \pi_{lj} \nonumber \\ 
& & + \frac{1}{N^2} \pi_{kl} \, \pi^k_j - \frac{1}{N} \p_s \pi_{jl} + \Omega_{jl} \, , 
% \nonumber \\ & & + \frac{1}{N^2} f_j n^\l \p_l N_\l - \frac{1}{N^2} Q_{\l \nu} \Bigl[ e^\nu_l \p_j n^\l - 
% \p_s (e^\l_j e^\nu_l) \Bigr] + \frac{N}{2} \pi_{kj} A^k G_{\l \nu} \p_l (n^\l n^\nu) \nonumber \\
% & & + \frac{\pi_{jk}}{2N} (g^{pk} + N^2 A^p A^k)\biggl\{ 
% \frac{N^\l}{N^2} (g_{kl} \p_p E^k_\l + g_{kp} \p_l E^k_\l + A_l \p_p N_\l + A_p \p_l N_\l) \nonumber \\
% & & \hskip100pt 
% + G_{\l \nu} \Bigl[ \p_p (n^\l e^\nu_l) - \p_l (n^\l e^\nu_p) - A^k \p_k (e^\l_l e^\nu_p) \Bigr] \biggr\}  
\e 
where $\Omega_{jl}$ contains only terms which are proportional to derivatives of the basis vectors and their 
duals with respect to $(y^k, s)$. \\
The five-dimensional Ricci tensor can easily be calculated from (\ref{Riemann}):
\b
R_{\mu \nu} & = & E^j_\mu E^l_\nu \Bigl[ h^{ik} U_{ijkl} 
- N^2 A^k (V_{jkl} + V_{lkj}) + N^2 W_{jl} \Bigr] \nonumber \\
& & + (N_\mu E^l_\nu + N_\nu E^l_\mu) (- h^{jk} V_{jkl} +  N^2 A^j W_{jl})  \nonumber \\
& & + N_\mu N_\nu h^{jl} W_{jl} \, .
\e
Then the five-dimensional Einstein's equations in vacuum 
\b
\label{5}
R_{\mu \nu} = 0
\e
reduce to
\b
\label{1}
h^{ik} U_{ijkl} - N^2 A^k (V_{jkl} + V_{lkj}) + N^2 W_{jl} & = &  0 \, , \\
\label{2}
h^{jk} V_{jkl} - N^2 A^j W_{jl} & = & 0  \, , \\
\label{3}
h^{jl} W_{jl} & = & 0 \, . 
\e
Multiplying (\ref{1}) by $A^j$ and adding it to (\ref{2}) allows us to exclude $W_{jl}$ from equation 
(\ref{2}). Then, using the expressions (\ref{u}) and (\ref{v}) for $U_{ijkl}$ and $V_{jkl}$, (\ref{2}) 
becomes:
\b
\label{max}
\nabla_k \, \pi^k_l - \nabla_l \, \pi^k_k = 0 \, ,
\e
Equations (\ref{max}) (as we will see below) are a {\it generalisation} of Maxwell's equations in a fixed gauge. \\
One has to make a very important point here. Klein's theory corresponds to a threading decomposition of the five-dimensional 
spacetime \cite{boersma}. Rigorous analysis \cite{zel} shows that the curvature tensor of the hypersurface formed is given by Zelmanov's 
curvature tensor, which differs from the ordinary Riemann curvature tensor by additional terms containing $s$-derivatives 
of the four-dimensional metric. The cylinder condition forces the two curvature tensors equal and thus represents a 
surface forming condition. In Kaluza's theory, the foliation of the five-dimensional spacetime corresponds to slicing \cite{boersma}.
Then the four-dimensional metric $g_{ij}$ naturally appears as the slicing metric and imposing a cylinder condition is not at all
necessary. \\
To simplify the analysis of the physics described by the fields $A_i$, $\f$, and $N$, we will, however, put aside the $s$-dependent terms.
Also for simplicity, we will assume that the basis elements and their duals are constant (thus recovering the original Kaluza's theory). 
The tensor $\Omega_{jl}$ will then vanish from (\ref{tuka}). \\
Equation (\ref{max}) becomes:
\b
\label{maxgen}
\nabla_k F^{kl} = - 2 A_k r^{kl} + \frac{2}{N^2} (\pi^{kl} - \pi^j_j g^{kl}) \p_k N \, .
\e
Here $\, F_{kl} = \nabla_k A_l - \nabla_l A_k \,$ is the Maxwell electromagnetic tensor with $A_k$ 
being the electromagnetic potential. \\
The first term on the right-hand-side of (\ref{maxgen}) describes an interaction between electromagnetic and
gravitational fields. We assume that it is much smaller than the remaining terms, so that we can neglect it. Note that $A_k$ cannot 
be ``gauged up'' to increase the scale of $A_k r^{kl}$. Furthermore, if $N$ is a constant, then (\ref{maxgen}) becomes the usual 
Maxwell's equations
\b
\label{maxus} 
\nabla_k F^{kl} = 0 \, .
\e
The remaining two equations are:
\b
\label{div}
\nabla_k (\frac{f^k}{N}) & = & 0 \, , \\
\label{einstein}
r_{jl} - \frac{1}{2}g_{jl}r & = & \frac{N^2}{2} T_{jl} \, ,
\e
where $r = g^{ik} r_{ik}$ and $r_{jl}$ are the four-dimensional scalar curvature and four-dimensional Ricci 
tensor, respectively. \\
% One can show that:
% \b
% r & = & (\pi^k_k)^2 - \pi^{ki} \pi_{ki} \nonumber \\
% & & \hskip30pt  = N \nabla_i (\pi^{ik} A_k - \pi^k_k A^i) \, , \\
% r_{jl} & = & \frac{2}{N^2} (\p_j N)(\p_l N) - \frac{1}{N} \nabla_j \nabla_l N 
% - 2 \pi_{kl} \pi^k_j \nonumber\\
% & & \hskip10pt - N \Bigl[\pi_{kj} \nabla_l A^k + \pi_{kl} \nabla_{j} A^k + \nabla_k (A^k \pi_{jl}) \Bigr]
% \e
The energy-momentum tensor $T_{jl}$ is therefore given by:
\b
T_{jl} = T^{\mbox{\tiny Maxwell}}_{jl} \, + \, g ^{ik} \nabla_i B_{jlk} \, + \, C_{jl} \, + \, D_{jl} \, ,  
\e
where:
\b
T^{\mbox{\tiny Maxwell}}_{jl} & = & g^{ik} F_{ij} F_{kl} - \frac{1}{4} g_{jl} F_{ik} F^{ik} \, , 
\nonumber \\ 
B_{jlk} & = & A_k \nabla_l A_j - A_l \nabla_k A_j - A_j F_{kl} + \nabla_j (A_k A_l) \nonumber \\
& & + g_{jl}(A^i \nabla_k A_i - A_k \nabla_i A^i)  \, ,  \\ 
C_{jl} & = & g_{jl} A^i A^k r_{ik} - 2 A^i A_l r_{ij} - 2 A^i A_j r_{il} \, , \\
D_{jl} & = & \frac{4}{N^4} (\p_j N)(\p_l N) \nonumber \\
& & - \frac{2}{N^3} \nabla_j \nabla_l N - \frac{2}{N^2} \pi^k_k (A_l \p_j N + A_j \p_l N) \nonumber \\
& & + \frac{2}{N^2} \Bigl[ - A^k \pi_{jl} + A_l \pi^k_j + A_j \pi^k_l \nonumber \\ 
& & \hskip48pt - g_{jl} (A^i \pi^k_i - A^k \pi^i_i)\Bigr] \p_k N \, .
\e
We will analyse each of these terms separately. The first one, $T^{\mbox{\tiny Maxwell}}_{jl}$, is the 
Maxwell energy-momentum tensor. The tensor $C_{jl}$ describes interaction between electromagnetic and 
gravitational fields. From (\ref{einstein}) we see that, if $N^2$ is very small (as we will confirm later), 
then $r_{jl}$ will be of the order of $N^2$, which justifies the neglection of the interaction terms in 
(\ref{maxgen}) and the tensor $C_{jl}$. \\
Using (\ref{KK}) in (\ref{f}) and then (\ref{f}) in equation (\ref{div}), we see that a constant solution 
for $N$ is allowed by equation (\ref{div}) if $\f$ satisfies:
\b
\label{fi}
\nabla^k \p_k \, \f = \frac{1}{2} F^{ik} F_{ik} \, ,
\e
where $F^{ik}$ is a solution of (\ref{maxus}). For the constant solution for $N$, the tensor $D_{jl}$ 
vanishes. Moreover:
\b
\label{conserv}
0 \equiv g^{ij} \nabla_i T_{jl} = g^{ij}\nabla_i T^{\mbox{\tiny Maxwell}}_{jl} \, , 
\e  
since $g^{ml} g^{nk} \nabla_m \nabla_n B_{jlk} = - \frac{2}{N} \nabla_j \nabla_k (\frac{f^k}{N}) = 0$
in view of (\ref{div}). \\
In other words, the conservation law (\ref{conserv}) is given by the usual Maxwell energy-momentum tensor
$T^{\mbox{\tiny Maxwell}}_{jl}$ and $N^2$ plays the role of the Newton's constant $G_N$:
\b
\label{nc}
\frac{N^2}{2} = \frac{8 \pi G_N}{c^4} \, .
\e
One has to point out here that in the set-up of Thiry \cite{th}, and in \cite{emb3}, $G_N \simeq \f^2$, 
where $\f$ satisfies (\ref{fi}). This implies that a constant solution for $\phi$ and, respectively, $G_N$ 
is only possible when the unphysical constraint $F^{ik} F_{ik} = 0$ is satisfied. \\
In contrast, in the dual set-up, a constant solution is possible. However, $N$ (together with $A_i$ and $g_{ij}$) 
is a solution to the system of equations (\ref{maxgen}), (\ref{div}), and (\ref{einstein}) and, in general, 
does not need to be a constant. Then it plays the role of a dilaton field. \\
To illustrate this, consider the standard cosmological metric \cite{scm} with $E^\nu_\mu = \delta^\nu_\mu$: 
\b
ds_{(5)}^2 & = & - s^2 dt^2 + t^{2/ \a} s^{2/(1-\a)} (dr^2 + r^2 d\Omega^2) \nonumber \\
&& \hskip60pt +\a^2(1 - \a)^{-2} t^2 ds^2 \, .
\e
Changing variables by: $r \to s^\gamma e^{\beta r}$ with $\gamma = - (1/2) (1+\a)/(1-\a)$ and $t^{1/\a} \to a(t)$, we get:
\b
ds_{(5)}^2 & = & - s^2 \a^2 [a(t)]^{2 \a - 2}\dot{a}^2(t) dt^2 + 2 A_r drds \nonumber \\
& & \, + \, s a^2(t) e^{2 \beta r} (\beta^2 dr^2 + d\Omega^2) + \f ds^2 \, ,
\e
where $A_r = \gamma \beta a^2(t) e^{2 \beta r}$, $A_t = A_\varphi =  A_\theta = 0$, and
$\f = (\gamma^2 / s) a^2(t) e^{2 \beta r} + \a^2(1 - \a)^{-2} a^{2 \a}(t)$. We take $\beta$ to be a negative constant, so that the field 
$A_r$ will fall off towards infinity. Since $a(t)$ describes the inflation, we note that the field $A_r$ expands as $a^2(t)$. The dilaton 
(which models the Newton's constant) varies as:
\b
N^2 = (1 - \a)^2 \a^{-2} [a(t)]^{-2 \a} \, .
\e
Thus:
\b
\frac{\dot{G}}{G} = - 2 \a \frac{\dot{a}}{a} = - 2 \a H \, ,
\e
where $H$ is the Hubble's constant. Observational limits \cite{gib} put $\a < 10^{-3}$. \\
One should note that the four-dimensional metric is now $s$-dependent, but this does not pose a problem in the slicing
formulation. Only the term $\p_s g_{jl}$ from the extrinsic curvature (\ref{KK}) should be recovered. \\
Finally, we note that the general solution for the dilaton field can be written as:
\b
N^2 = \frac{(\det g_{ik}) (\det E^\mu_\nu)^2}{\det G_{\mu \nu}} \, , 
\e
for a solution $G_{\mu \nu}$ of (\ref{5}) and embedding specified with $E^\mu_\nu$. \\
To recapitulate, we have found plausable generalisations of Einstein--Maxwell equations and explained the 
origin of the constant solution for the dilaton (representing the Newton's constant $G_N$) as well as the 
possibilities for modelling non-constant solutions for different cosmologies (representing time-varying $G_N$) in relation
to the gauge freedom of our model.  

\vskip.4cm
\noindent
We thank Vesselin Gueorguiev, Brian Dolan, Brien Nolan, Siddhartha Sen and the anonymous referee for useful 
discussions and comments. \\
E.M.P. dedicates this work to his son Emanuel.

\end{document}